\documentclass[aps,prb,reprint,twocolumn,notitlepage,superscriptaddress,longbibliography]{revtex4-1}
\usepackage{graphicx}
\usepackage{amsmath,amssymb}
\usepackage{color}
\usepackage{bm}
\usepackage{bbm}
\usepackage{mathtools}
\usepackage{hyperref}

\begin{document}

\title{Quadratic dispersion relations in gapless frustration-free systems}
\author{Rintaro Masaoka}
\affiliation{Department of Applied Physics, The University of Tokyo, Tokyo 113-8656, Japan}
\author{Tomohiro Soejima
}
\affiliation{Department of Physics, Harvard University, Cambridge, MA 02138, USA}
\author{Haruki Watanabe}\email{hwatanabe@g.ecc.u-tokyo.ac.jp}
\affiliation{Department of Applied Physics, The University of Tokyo, Tokyo 113-8656, Japan}
\date{\today}

\begin{abstract}
Recent case-by-case studies revealed that the dispersion of low energy excitations in gapless frustration-free Hamiltonians is often quadratic or softer. In this worl, we argue that this is actually a general property of such systems. By combining the previous study by Bravyi and Gosset and the min-max principle, we prove this hypothesis for models with local Hilbert spaces of dimension two that contains only nearest-neighbor interactions on cubic lattice. This may be understood as a no-go theorem realizing gapless phases with linearly dispersive excitations in frustration-free Hamiltonians.  We also provide examples of frustration-free Hamiltonians in which the plane-wave state of a single spin flip does not constitute low energy excitations.
\end{abstract}
\maketitle

\section{Introduction}
Accurately investigating the ground states and excited states of quantum many-body systems is often challenging. It is hence common practice to analyze the system of interest using the simplest possible model that belongs to the same phase. Frustration-free (FF) systems, in particular, are relatively easy to handle, and their exact ground states and excited states are sometimes accessible. For these reasons, FF systems have been widely used as representative models of various phases of matter. Therefore, elucidating the general properties of FF systems and understanding the limits of the phases that can be realized by FF Hamiltonians hold significant importance.

Numerous FF systems possess an excitation gap. Notable examples include the Majumdar--Ghosh (MG) model, known as a toy model realizing spontaneous breaking of discrete translational symmetry, the Affleck--Kennedy--Lieb--Tasaki model, representing a symmetry-protected topological phase, and the Kitaev toric code model, serving as the canonical model for topologically ordered phases~\cite{Auerbach,TasakiBook}. The parent Hamiltonians of Matrix Product States (MPS) also feature an excitation gap~\cite{parent}.

There are also FF Hamiltonians with gapless excitations. For example, the uncle Hamiltonian of MPS~\cite{uncle}, various models at critical points, and models related to spontaneously broken continuous symmetries~\cite{PhysRevLett.131.220403,arXiv:2310.16881} are gapless. Through their recent investigations, the general properties of FF systems have gradually become clearer.

In this worl, we discuss two conjectures regarding FF systems. The first one is about the finite-size splitting between degenerate ground states. Generally, even if the ground states are degenerate in the thermodynamic limit, their energy eigenvalues in a finite system do not match due to the finite-size splitting. However, in FF systems, the ground states are exactly degenerate even before taking the thermodynamic limit. In other words, whether a frusration-free system is gapped or not can be judged by looking at the limiting behavior of the difference between the ground state energy and the second lowest energy level. 

The second conjecture concerns the dispersion relation of low-energy excitations in FF systems. Although gapless modes typically have a linear dispersion, it is known that the ferromagnetic Heisenberg model, a representative FF Hamiltonian, have a spin wave excitation with a quadratic dispersion $E(k)=2\sin^2(k/2)$ \cite{Auerbach}. In fact, our second conjecture asserts that gapless FF systems always have a quadratic or softer dispersion relation. Although there are several arguments supporting this statement in literature~\cite{PhysRevB.103.214428,PhysRevLett.131.220403,arXiv:2310.16881,arXiv:2405.00785}, they are applicable only to the case where the low-energy excitations are obtained by applying a sum of local operators to a ground state.
We provide a general proof for spin-1/2 models with nearest-neighbor interaction on cubic lattice based on Ref.~\onlinecite{BravyiGosset}. We also discuss examples with longer-range interactions and show that gapless excitations with quadratic dispersion may not be obtained by a sum of local operators.

\section{Overview}
\subsection{Setting and definitions}
In this work, we consider Hamiltonians 
\begin{align}
\hat{H}=\sum_{\bm{r}\in\Lambda}\hat{H}_{\bm{r}}
\end{align}
defined on a finite $d$-dimensional lattice $\Lambda$.
The local Hilbert space on each lattice site $\bm{r}\in\Lambda$ is finite dimensional, which is denoted by $D_0$.  
Let $|\Lambda|$ be the number of lattice sites in $\Lambda$. Then the dimension of the total Hilbert space is $D=D_0^{|\Lambda|}$.
We assume that the Hamiltonian is local in the sense that $\hat{H}_{\bm{r}}$ acts nontrivially on the sites within a finite distance $R$ from $\bm{r}$.  The Hamiltonian is translation invariant if $\hat{T}_{\bm{a}}\hat{H}_{\bm{r}}=\hat{H}_{\bm{r}+\bm{a}}\hat{T}_{\bm{a}}$ for each lattice vector $\bm{a}$ and $\bm{r}\in\Lambda$. This is possible only when the periodic boundary condition (PBC) is imposed.

Let us write eigenvalues of $\hat{H}$ as 
\begin{align}
E_1\leq E_2\leq\cdots\leq E_D.
\end{align}
The thermodynamic limit $|\Lambda|\to\infty$ is taken by a sequence of increasing system size. The system is said to be gapped if there exists an integer $N_{\mathrm{deg}}$ ($1\leq N_{\mathrm{deg}}< D$), which may depend on the system size, such that $\lim_{|\Lambda|\to\infty}\big(E_{N_{\mathrm{deg}}}-E_1\big)=0$ and $\lim_{|\Lambda|\to\infty}\big(E_{N_{\mathrm{deg}}+1}-E_{N_{\mathrm{deg}}}\big)\neq0$. Otherwise the system is gapless, and the spectrum above the ground state is continuous.
When $N_{\mathrm{deg}}\geq2$, the quantity $E_{N_{\mathrm{deg}}}-E_1$ is referred to as the finite-size splitting among degenerate ground states. 
When the degeneracy originates from spontaneous breaking of a discrete symmetry or formation of a topological order, it generically decays exponentially with the system size. 
Even when $N_{\mathrm{deg}}$ converges to a finite number in the large $|\Lambda|$ limit, its value for a finite system may be ambiguous due to the possible system-size dependence.

The Hamiltonian is FF if a ground state $|{\Phi_0}\rangle$ of $\hat{H}$ is a simultaneous ground state of every local Hamiltonian $\hat{H}_{\bm{r}}$~\cite{TasakiBook}.  In this case, without loss of generality nor modifying the ground states and low-energy excitations of the system, one may assume that each $\hat{H}_{\bm{r}}$ is a projector, i.e., 
$\hat{Q}_{\bm{r}}^2=\hat{Q}_{\bm{r}}$ and $\hat{Q}_{\bm{r}}|{\Phi_0}\rangle=0$ for all $\bm{r}\in\Lambda$. Under this choice, the ground state energy becomes $E_1=0$. 

\subsection{Conjectures}
Let us state our conjectures one by one. 
\subsubsection{Conjecture 1}
The first conjecture is that, when the Hamiltonian is FF and gapped, the finite-size splitting among ground states precisely vanishes before taking the thermodynamic limit; i.e., $E_{N_{\mathrm{deg}}}=E_1$. For example, in the AKLT model for spin $s=1$ chain under OBC, the four-fold ground state degeneracy  due to the edge modes is exact. The four-fold topological degeneracy in the Kitaev toric code under PBC is also exact. This observation implies the following criterion for the excitation gap in FF systems. Let $\tilde{N}_{\mathrm{deg}}$ ($1\leq \tilde{N}_{\mathrm{deg}}< D$) be the number of exact zero energy states, i.e., $E_{\tilde{N}_{\mathrm{deg}}}=0$ and $E_{\tilde{N}_{\mathrm{deg}}+1}\neq 0$, which may also depend on the system size.
The Hamiltonian is gapless if and only if 
\begin{align}
\lim_{|\Lambda|\to\infty}E_{\tilde{N}_{\mathrm{deg}}+1}=0.
\end{align}
Actually this criterion has been used as the definition of the excitation gap in previous studies~\cite{BravyiGosset,Gossetozgunov,GossetHuang}.
Note that $\tilde{N}_{\mathrm{deg}}$ is well-defined even for gapless systems unlike $N_{\mathrm{deg}}$. 

\subsubsection{Conjecture 2}
The second conjecture is that, when the Hamiltonian is translation-invariant, FF, and gapless, there exists a family of variational states $|\Psi_{\bm{k}}\rangle$ that are orthogonal to all the ground states, are eigenstates of translation operator $\hat{T}_{\bm{a}}|\Psi_{\bm{k}}\rangle=e^{-i\bm{k}\cdot\bm{a}}|\Psi_{\bm{k}}\rangle$, and have a quadratic dispersion about the gapless point at $\bm{k}=\bm{k}_0$:
\begin{align}
\langle\Psi_{\bm{k}}|\hat{H}|\Psi_{\bm{k}}\rangle=O\big(|\bm{k}-\bm{k}_0|^2, L^{-2}\big)\label{prop2}
\end{align}
For systems under OBC, the result of Refs.~\onlinecite{Knabe,Anshu,Gossetozgunov,Lemm_2022} gives us the bound $E_{\tilde{N}_{\mathrm{deg}}+1}= O(L^{-2})$. Although this is consistent with our conjecture, it is not directly applicable to the PBC case~\cite{arXiv:2310.16881}.

In Sec.~\ref{sec1D}, we give a proof of this conjecture for $s=1/2$ spin chain with nearest neighbor interactions.
We further generalize the results to higher dimensions in Sec.~\ref{sec2D}.

\subsection{Min-max principle}
Our discussions below are based on a mathematical theorem called the min-max principle (see, theorem A.7 of Ref.~\onlinecite{TasakiBook}).
We compare two Hamiltonians $\hat{H}$ and $\hat{H}'$ acting on the same $D$-dimensional Hilbert space.  The eigenvalues of $\hat{H}$ and $\hat{H}'$ are, respectively, denoted by $E_j$ and $E_j'$ ($j=1,2,\cdots,D$) in the increasing order. Suppose that $\hat{V}\coloneqq\hat{H}'-\hat{H}$ is positive semi-definite $\hat{V}\geq0$. That is, $\langle\Phi|\hat{V}|\Phi\rangle\geq0$ for any state $|\Phi\rangle$.  In this setting, the min-max principle states that 
\begin{align}
E_j'\geq E_j\quad ({}^\forall j=1,2,\cdots,D)
\end{align}
For readers' convenience, we review the proof in Appendix~\ref{appminimax}. This is trivial when the two Hamiltonians can be diagonalized by a common unitary operator, but it holds more generally. 

Two corollaries follow immediately from the theorem. Let us consider two FF Hamiltonians $\hat{H}=\sum_{\bm{r}\in\Lambda}\hat{Q}_{\bm{r}}$ and $\hat{H}'=\sum_{\bm{r}\in\Lambda}\hat{Q}_{\bm{r}}'$. Suppose that $\hat{V}=\hat{H}'-\hat{H}$ is positive-semidefinite.
Then the minimax principle implies that the number of zero-energy states of $\hat{H}$ cannot be smaller than that of $\hat{H}'$: $\tilde{N}_{\mathrm{deg}}\geq \tilde{N}_{\mathrm{deg}}'$. Furthermore, if (i) $\hat{H}$ is gapped and (ii) $\tilde{N}_{\mathrm{deg}}$ is bounded by a system-size independent constant, then $\hat{H}'$ is also gapped. This is because, if $\hat{H}'$ were gapless, an increasing number of eigenstates of $\hat{H}'$ have lower energy than
$E_{\tilde{N}_{\mathrm{deg}}+1}>0$ and eventually $E_{\tilde{N}_{\mathrm{deg}}+1}>E_{\tilde{N}_{\mathrm{deg}}+1}'$ holds for a sufficiently large $L$. This relation was previously used for interacting Kitaev chains~\cite{PhysRevB.92.115137,PhysRevB.98.155119}.

\section{1D system with nearest-neighbor interactions}
\label{sec1D}
In this section, we discuss a chain of $L$ qubits following Ref.~\onlinecite{BravyiGosset}.
The Hilbert space on each qubit is spanned by $|0\rangle$ and $|1\rangle$, which may be interpreted as $|{\uparrow}\rangle$ and $|{\downarrow}\rangle$ for an $s=1/2$ spin. 
Let us consider the Hamiltonian $\hat{H}\coloneqq\sum_{x=1}^L\hat{Q}_{x,x+1}$ with nearest-neighbor interactions under PBC.  Here, $\hat{Q}_{x,x+1}=\hat{Q}_{x,x+1}^2$ is a projector nontrivially acting on the spins at $x$ and $x+1$.
The number of $+1$ eigenvalues of a projector is called the rank. The Hamiltonian $\hat{H}$ has the translation symmetry $\hat{T}_1$ which satisfies $\hat{T}_1\hat{Q}_{x-1,x}=\hat{Q}_{x,x+1}\hat{T}_1$. 

\begin{figure*}[t]
\begin{center}
\includegraphics[width=\textwidth]{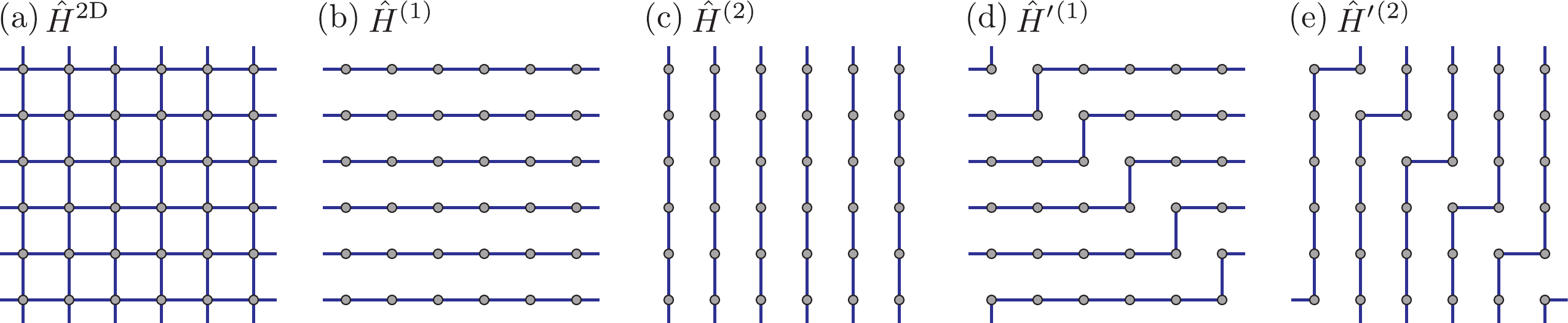}
\end{center}
\caption{The illustration of the 2D model and its decomposition into 1D models.
\label{fig2dlattice}
}
\end{figure*}

The seminal work by Bravyi and Gosset~\cite{BravyiGosset} showed that the Hamiltonian $\hat{H}^{\mathrm{obc}}\coloneqq\sum_{x=1}^{L-1}\hat{Q}_{x,x+1}$ under open boundary condition (OBC) can be gapless only when the projector $\hat{Q}_{x,x+1}$ is unitary equivalent to Eqs.~\eqref{BGHamiltonian2}, \eqref{BGHamiltonian3} for the rank 1 case and Eq.~\eqref{Qf} for the rank 2 case. 
Then the min-max principle implies the same for $\hat{H}$ under PBC, since $\hat{H}-\hat{H}^{\mathrm{obc}}=\hat{Q}_{L,1}\geq0$ is positive semi-definite.
We will see that the fully polarized state $|{\Phi_0}\rangle\coloneqq\bigotimes_{x=1}^{L}|0\rangle_x=|0\cdots0\rangle$ is a common ground state of these Hamiltonians. Furthermore, the plane-wave state
\begin{align}
&|{\Psi_k}\rangle\coloneqq\frac{1}{\sqrt{L}}\sum_{x=1}^Le^{ikx}\hat{s}_x^-|{\Phi_0}\rangle\label{BGvs}
\end{align}
with $k\coloneqq2\pi m/L$ ($m=0,1,2,\cdots L-1$) is a common variational state with a quadratic dispersion relation.  

\subsection{Rank 1 case}
Let us start with the rank 1 case. We write
\begin{align}
\hat{Q}_{x,x+1}\coloneqq|\psi\rangle_{x,x+1}\langle\psi|_{x,x+1},\label{BGHamiltonian2}
\end{align}
where $|\psi\rangle_{x,x+1}$ is a normalized state for two spins at $x$ and $x+1$.  Without loss of generality one can assume~\cite{BravyiGosset}
\begin{align}
&|\psi\rangle=(\alpha+i\beta)|01\rangle+(\alpha+i\gamma)|10\rangle+\delta |11\rangle\label{BGHamiltonian3}
\end{align}
with $\alpha, \beta, \gamma,\delta\in\mathbb{R}$ and $2\alpha^2+\beta^2+\gamma^2+\delta^2=1$, 
by performing a local unitary transformation (see Appendix~\ref{derivationpsi}).  For example, when $(\alpha,\beta,\gamma,\delta)=\frac{\pm1}{\sqrt{2}}(0,1,-1,0)$, $\hat{Q}_{x,x+1}=\hat{\Pi}_{x,x+1}^{s=0}\coloneqq \frac{1}{4}\hat{\mathbbm{1}}-\hat{\bm{s}}_x\cdot\hat{\bm{s}}_{x+1}$ is the projector onto the spin-singlet state $|s\rangle\coloneqq(|01\rangle-|10\rangle)/\sqrt{2}$ and the corresponding Hamiltonian is the ferromagnetic Heisenberg model.
Any state that is invariant under permutations of any two qubits (e.g., the fully-polarized state and the W state) is a ground state of this model.
As a basis of such states, we can choose $\{\hat{\mathcal{S}}|n\rangle\}_{n=0}^L$, where $|n\rangle\coloneqq |\underbrace{0\cdots0}_{L-n}\underbrace{1\cdots1}_{n}\rangle$ and $\hat{\mathcal{S}}$ is the symmetrization operator, uniformly averaging over the $\binom{L}{n}=\frac{L!}{(L-n)!n!}$ states.  

To analyze more general cases,  let us introduce a matrix
$m_\psi\coloneqq\begin{pmatrix}
\alpha-i\beta&\delta\\
0&-\alpha+i\gamma
\end{pmatrix}$.
The model defined by Eqs.~\eqref{BGHamiltonian2}, \eqref{BGHamiltonian3} can be gapless only when $\det m_\psi=(\alpha-i\beta)(-\alpha+i\gamma)\neq0$, i.e., $|\psi\rangle$ is entangled~\cite{BravyiGosset}.  
In this case, we introduce an invertible operator $\hat{M}$ composed of nonuniform powers of the matrix $m_\psi$:
\begin{align}
&\hat{M}\coloneqq\mathbbm{1}\otimes\hat{m}_\psi\otimes\hat{m}_\psi^2\otimes\cdots\otimes\hat{m}_\psi^{L-1},\\
&\hat{m}_\psi\coloneqq(\alpha-i\beta)|0\rangle\langle0|-(\alpha-i\gamma)|1\rangle\langle1|+\delta|0\rangle\langle1|.
\end{align}
This operator maps $|\psi\rangle$ to the singlet state $\hat{M}^\dagger|\psi\rangle=(\det{m_\psi}^*)^{j}|s\rangle$. It follows that $\hat{Q}_{x,x+1}$ for the state in Eq.~\eqref{BGHamiltonian3} can be connected to $\hat{\Pi}_{x,x+1}^{s=0}$ as
\begin{align}
\hat{Q}_{x,x+1}=4|{\det{m}_\psi}|^2(\hat{M}\hat{\Pi}_{x,x+1}^{s=0}\hat{M}^{-1})^\dagger(\hat{M}\hat{\Pi}_{x,x+1}^{s=0}\hat{M}^{-1}).
\end{align}
This type of non-unitary transformation is known as Witten's conjugation~\cite{WITTEN1982253,katsura}. As a consequence, ground states of $\hat{H}^{\mathrm{obc}}$ takes the form $\hat{M}\hat{\mathcal{S}}|n\rangle$ ($n=0,1,\cdots L$)~\cite{BravyiGosset}.

The Hamiltonian under PBC has an additional term $\hat{Q}_{L,1}$. 
Every ground state of $\hat{H}^{\mathrm{obc}}$ remains a ground state of $\hat{H}$ if $m_\psi^L$ is proportional to the identity matrix $\mathbbm{1}$. However, except when  $(\alpha,\beta,\gamma,\delta)=\frac{\pm1}{\sqrt{2}}(0,1,-1,0)$ discussed above, $m_\psi^L$ is not proportional to $\mathbbm{1}$ for some $L$. In such a case, the ground state of $\hat{H}$ is restricted to the product state of the eigenstates of $\hat{m}_\psi$; that is, $|{\Phi_0}\rangle=|00\cdots 0\rangle$ and $|uu\cdots u\rangle$, where $|u\rangle\coloneqq\frac{[2\alpha-i(\beta+\gamma)]|1\rangle-\delta |0\rangle}{\sqrt{4\alpha^2+(\beta+\gamma)^2+\delta^2}}$.

The variational state $|{\Psi_k}\rangle$ in Eq.~\eqref{BGvs} with $k\neq0$ is orthogonal to the ground states at least when $m_\psi^L$ is not proportional to $\mathbbm{1}$, since their translation eigenvalues are different. The energy expectation value $\langle\Psi_{k}|\hat{H}|\Psi_{k}\rangle$ is given by
\begin{align}
E(k)&= \big|(\alpha-i\gamma) +(\alpha-i\beta) e^{ik}\big|^2\notag\\
&=(1-\delta^2)-A\cos (k-k_0),
\end{align}
where $A\coloneqq\sqrt{(1-\delta^2)^2-(\beta^2-\gamma^2)^2}$ and $k_0$ is defined by $\cos k_0=-2(\alpha^2+\gamma\beta)/A$ and $\sin k_0=-2\alpha(\beta-\gamma)/A$.
This excitation energy takes the minimum value at $k=k_0$, which vanishes if $\beta^2=\gamma^2$, implying the existence of gapless excitations.
In this case $E(k)$ can be rewritten as $E(k)=2(1-\delta^2)\sin^2((k-k_0)/2)$, confirming our conjecture in this model.  When $\gamma=\beta$, $k_0$ is precisely $\pi$, but it takes more general value when $\gamma=-\beta$.

\subsection{Rank 2 case}
Next we discuss the rank 2 case. This time the Hamiltonian can be gapless only when $\hat{Q}_{x,x+1}$ is unitary equivalent to the XY model with an external magnetic field~\cite{BravyiGosset}:
\begin{align}
&\hat{Q}_{x,x+1}\coloneqq
\frac{1}{2}\hat{\mathbbm{1}}-\frac{\zeta \hat{s}_x^+\hat{s}_{x+1}^-+\text{h.c.}}{1+|\zeta|^2}-\frac{|\zeta|^2\hat{s}_x^z+\hat{s}_{x+1}^z}{{1+|\zeta|^2}}\label{Qf}
\end{align}
with $\zeta\in\mathbb{C}$. This Hamiltonian can also be expressed as $\mathrm{ker}\,\hat{Q}_{x,x+1}=\mathrm{span}\big\{|00\rangle,|10\rangle+\zeta|01\rangle\big\}$.
When $\zeta=1$, ground states of $\hat{Q}_{x,x+1}^{\zeta=1}$ are the fully-polarized state $|{\Phi_0}\rangle$ and the zero momentum state of single spin flip $|\Psi_0\rangle$, regardless of the boundary condition. 
We find that $\hat{Q}_{x,x+1}$ with $\zeta\neq1$ can be obtained from $\hat{Q}_{x,x+1}^{\zeta=1}$ by Witten's conjugation as
\begin{align}
\hat{Q}_{x,x+1}=(\hat{M}\hat{Q}_{x,x+1}^{\zeta=1}\hat{M}^{-1})^\dagger\hat{C}(\hat{M}\hat{Q}_{x,x+1}^{\zeta=1}\hat{M}^{-1}),
\end{align}
where $\hat{M}\coloneqq\hat{\mathbbm{1}}\otimes \hat{m}_\zeta\otimes\cdots\otimes \hat{m}_\zeta^{L-1}$, $\hat{m}_\zeta\coloneqq|0\rangle\langle0|+\zeta|1\rangle\langle1|$, and $\hat{C}\coloneqq\hat{\mathbbm{1}}+\frac{(1-|\zeta|^2)^2}{2\mathrm{Re}\zeta(1+|\zeta|^2)}(|10\rangle\langle01|+|01\rangle\langle10|)$.
Hence, the ground states of $\hat{H}^{\mathrm{obc}}$ with $\zeta\neq1$ are given by $\hat{M}|{\Phi_0}\rangle=|{\Phi_0}\rangle$ and $\hat{M}|\Psi_0\rangle\propto\sum_{x=1}^L\zeta^{x-1}|x\rangle.$
The latter remains a ground state under PBC if and only if $\zeta^L=1$.

Because of  the U(1) symmetry of the Hamiltonian in Eq.~\eqref{Qf}, the plane-wave state $|{\Psi_k}\rangle$ in Eq.~\eqref{BGvs} is actually an exact eigenstate of $\hat{H}$. The eigenenergy is
\begin{align}
E(k)=\frac{|e^{ik}-\zeta|^2}{1+|\zeta|^2}=\frac{1-|\zeta|^2}{1+|\zeta|^2}+\frac{4|\zeta|}{1+|\zeta|^2}\sin^2\Big(\frac{k-k_0}{2}\Big),
\end{align}
where $k_0$ is defined by $\zeta=|\zeta|e^{ik_0}$.
Hence, the model is gapless and the conjecture holds when $|\zeta|=1$.

Finally, for the gapped cases, we found that the ground state degeneracy is, in general, bounded as $N_{\mathrm{deg}}\leq2$, 
except the rank 1 model with $\alpha=\beta=\gamma=0$ for which 
\begin{align}
N_{\mathrm{deg}}=\sum_{n=0}^{\lfloor \frac{L+1}{2} \rfloor}\binom{L-n+1}{n}-\sum_{n=2}^{\lfloor \frac{L+1}{2} \rfloor}\binom{L-n-1}{n-2}.\label{Lucus}
\end{align}
See Appendix~\ref{secLucus} for the derivation. This coincides with the Lucas number~\cite{Lucas} that increases exponentially with the system size $L$. 

\section{2D models with nearest-neighbor interactions}
\label{sec2D}
In this section, we generalize results obtained in Sec.~\ref{sec1D} to higher dimensions. We consider a square lattice of $s=1/2$ spins, assuming the Hamiltonian of the form $\hat{H}\coloneqq\sum_{x,y=1}^{L}\big(\hat{Q}_{(x,y),(x+1,y)}+\hat{Q}_{(x,y),(x,y+1)}\big)$, 
where $\hat{Q}_{(x,y),(x',y')}$ is a projector acting on two spins at $(x,y)$ and $(x',y')$ [Fig.~\ref{fig2dlattice} (a)].  We impose PBC and assume the translation invariance in both $x$ and $y$ directions. Our goal is to show that, if $\hat{H}$ is FF and gapless, there exists a variational state $|\Psi_{\bm{k}}\rangle$ with the dispersion satisfying \eqref{prop2}.

To this end, let us decompose the 2D Hamiltonian into 1D chains in two different ways.
The first choice is the simple one, decomposing $\hat{H}$ into decoupled chains $\hat{H}^{(1)}\coloneqq\sum_{y=1}^{L}\hat{H}_y^{\mathrm{1D}}$ and $\hat{H}^{(2)}\coloneqq\sum_{x=1}^{L}\hat{H}_x^{\mathrm{1D}}$ as illustrated in Fig.~\ref{fig2dlattice} (b,c).
For each $y$, $\hat{H}_y^{\mathrm{1D}}\coloneqq\sum_{x=1}^{L}\hat{Q}_{(x,y),(x+1,y)}$ describes the Hamiltonian for the chain along the $x$ axis; 
similarly, for each $x$, $\hat{H}_x^{\mathrm{1D}}\coloneqq\sum_{y=1}^{L}\hat{Q}_{(x,y),(x,y+1)}$ describes the Hamiltonian for the chain along the $y$ axis. The second decomposition $\hat{H}=\hat{H}'{}^{(1)}+\hat{H}'{}^{(2)}$ is illustrated in Fig.~\ref{fig2dlattice} (d,e), in which the chains are connected into one piece.  Even when these 1D Hamiltonians are FF, whether the total Hamiltonian $\hat{H}$ remains FF or not is, in general, nontrivial. However, this issue can be easily gone around in our case. When either $\hat{H}'{}^{(1)}$ or $\hat{H}'{}^{(2)}$ is gapped with a finite ground state degeneracy, the min-max principle suggests that $\hat{H}$ is also gapped, because $\hat{H}'{}^{(2)}=\hat{H}-\hat{H}'{}^{(1)}$  and $\hat{H}'{}^{(1)}=\hat{H}-\hat{H}'{}^{(2)}$ are positive semi-definite~\footnote{This argument cannot be directly applied to the simpler decomposition $\hat{H}^{(1)}+\hat{H}^{(2)}$, because if $N_{\mathrm{deg}}\geq2$ for each $\hat{H}_y^{\mathrm{1D}}$, then the ground state degeneracy of $\hat{H}^{(1)}$ becomes $N_{\mathrm{deg}}^L$, which grows exponentially with the system size and the min-max principle becomes silent.
For the same reason, the case of $\alpha=\beta=\gamma=0$ in the rank 1 model should be treated separately, but this case is trivially gapped in any spatial dimension.
}.  Hence, we can restrict ourselves to the case where both $\hat{H}'{}^{(1)}$ and $\hat{H}'{}^{(2)}$ are gapless, implying that $\hat{H}^{(1)}$ and $\hat{H}^{(2)}$ are also gapless. 
Furthermore, as we have seen above, when the Hamiltonian for an $s=1/2$ spin chain with nearest-neighbor interaction is gapless, the fully polarized state $|{\Phi_0}\rangle$ is a common zero-energy ground state. Hence, the 2D version of the fully polarized state $|{\Phi_0}\rangle\coloneqq\bigotimes_{x,y=1}^{L}|0\rangle_{(x,y)}$ is a simultaneous ground state of all terms in $\hat{H}$, implying that  $\hat{H}$ is FF.

To construct a variational state for low-energy excitations, let us define the plane-wave state of a single spin flip by $|\Psi_{\bm{k}}\rangle\coloneqq\frac{1}{L}\sum_{x,y=1}^{L}e^{i\bm{k}\cdot\bm{r}}\hat{s}_{(x,y)}^-|{\Phi_0}\rangle$. The variational energy is given by
\begin{align}
E(\bm{k})&\coloneqq\langle\Psi_{\bm{k}}|\hat{H}|\Psi_{\bm{k}}\rangle=E(k_x)+E(k_y),\label{2dvariationenergy}
\end{align}
which satisfies \eqref{prop2}. This discussion can be readily extended to the cubic lattice. The variational energy of the plane-wave state $|\Psi_{\bm{k}}\rangle\coloneqq L^{-3/2}\sum_{x,y,z=1}^{L}e^{i\bm{k}\cdot\bm{r}}\hat{s}_{(x,y,z)}^-|{\Phi_0}\rangle$ is simply given by $E(k_x)+E(k_y)+E(k_z)$.

Finally, let us investigate possible generalization to other form of lattices.
For example, the nearest-neighbor interaction on the triangular lattice contains a diagonal interaction $\sum_{x,y=1}^{L}\hat{Q}_{(x,y),(x+1,y-1)}$. More generally, when an interaction $\hat{Q}_{(x,y),(x+d_x,y+d_y)}$ among the spins at $(x,y)$ and $(x+d_x,y+d_y)$ is added, the variational energy obtains a term $E(k_xd_x+k_yd_y)$. Hence, the plane-wave state generically becomes gapped unless $E(k_0(d_x+d_y))=0$. When $k_0=0$ (e.g., in the ferromagnetic Heisenberg model), this condition can be easily satisfied, but otherwise some  fine-tuning is required.

\section{Examples}
In this section we discuss several examples for which our general discussing in Sec.~\ref{sec1D} and \ref{sec2D} are not applicable. We summarize our results in Table~\ref{table1}.
\begin{table*}[t]
\centering
\caption{Comparison of gapless FF Hamiltonians for $s=1/2$ chain.
$|{\Phi_0}\rangle=|00\cdots\rangle$ is the fully polarized state and 
$|{\Psi_k}\rangle$ [Eq.~\eqref{BGvs}] is the plane-wave state of single spin flip. The XYZ MG model is discussed in Sec.~IV of SM.
}
\begin{tabular}{c|c|c|c|c}\hline\hline
Interaction type & Range & $|{\Phi_0}\rangle$ is a GS &  $|{\Psi_k}\rangle$ is gapless & Variational energy  \\\hline
Nearest neighbor models with rank 1: \eqref{BGHamiltonian2}, \eqref{BGHamiltonian3} with $\gamma=\pm\beta$ & 1 & \checkmark & \checkmark& $O\big((k-k_0)^2\big)$ \\
Nearest neighbor models with rank 2:  \eqref{Qf} with $|\zeta|=1$ & 1 & \checkmark & \checkmark& $O\big((k-k_0)^2\big)$\\
GHZ uncle Hamiltonian: \eqref{GHZ} & 2 & \checkmark & -- & $O\big(k^2,L^{-2}\big)$\\
XYZ MG model with $\theta\neq0$: \eqref{HXYZ} &  2  & -- & -- & $O\big(k^2)$\\\hline\hline
\end{tabular}
\label{table1}
\end{table*}

\subsection{Kinetic Ising model}
Let us consider an $s=1/2$ spin model defined on a $d$-dimensional cubic lattice: $\hat{H}=\sum_{\bm{r}}\hat{H}_{\bm{r}}$, where
\begin{align}
&\hat{H}_{\bm{r}}=\frac{1}{2\cosh(J\sum_{\bm{r}'\in B_{\bm{r}'}}\hat{\sigma}_{\bm{r}'}^z)}\Big(e^{-J\hat{\sigma}_{\bm{r}}^z{}\sum_{\bm{r}'\in B_{\bm{r}}}\hat{\sigma}_{\bm{r}'}^z}-\hat{\sigma}_{\bm{r}}^x\Big).
\end{align}
Here, $\hat{\sigma}_{\bm{r}}^a$ ($a=x,y,z$) are the Pauli matrices related to spin operators by $\hat{s}_{\bm{r}}^a=(1/2)\hat{\sigma}_{\bm{r}}^a$ and $B_{\bm{r}}$ is the nearest neighbors of $\bm{r}$.
This model is associated with the classical Ising model with the Boltzmann weight $\propto e^{J\sum_{\langle \bm{r},\bm{r}'\rangle}\sigma_{\bm{r}}^z\sigma_{\bm{r}'}^z}$~\cite{masaoka}. (The temperature $T$ is included in the definition of $J$.) The $\mathbb{Z}_2$ symmetry $\prod_{\bm{r}}\hat{\sigma}_{\bm{r}}^x$ of the model is spontaneously broken only in the limit $J\to\infty$, which corresponds to the zero temperature limit, in one dimension and for $J$ above a critical value $J_c$ in higher dimensions.

\subsubsection{1D}
In one dimension, $\hat{H}_x$ is a projector $\hat{Q}_{x-1,x,x+1}$ acting on the three spins at $x-1$, $x$, and $x+1$, which can be expressed as
\begin{align}
&\hat{Q}_{x-1,x,x+1}\notag\\
&=\frac{1}{2}\hat{\mathbbm{1}}-c_2\hat{s}_{x}^z(\hat{s}_{x-1}^z+\hat{s}_{x+1}^z)-2\hat{s}_{x}^x\Big(\frac{c_1}{4}-c_3\hat{s}_{x-1}^z\hat{s}_{x+1}^z\Big)\label{GHZ}
\end{align}
with $c_1=\frac{2\cosh^2J}{\cosh(2J)}$, $c_2=\tanh(2J)$, and $c_3=\frac{2\sinh^2J}{\cosh(2J)}$.

In the $J\to\infty$ limit, all three parameters $c_1$, $c_2$, $c_3$ become $1$ and the model reduces to the uncle Hamiltonian for the Greenberger--Horne--Zeilinger (GHZ) state~\cite{uncle}. The kernel of $\hat{Q}_{x,x+1,x+2}$  is given by
\begin{align}
\mathrm{span}\Big\{
|000\rangle,
|111\rangle,
|100\rangle+|110\rangle,
|001\rangle+|011\rangle\Big\}.
\end{align}

\begin{figure}[t]
\begin{center}
\includegraphics[width=1\columnwidth]{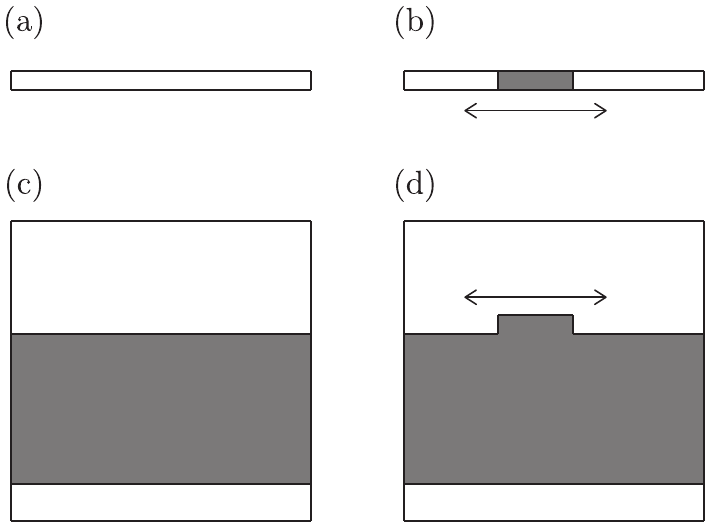}
\end{center}
\caption{
(a) The all spin-up state $|\Phi_0\rangle$ of 1D kinetic Ising model.
(b) An excited state $|\Psi_{k,\ell}\rangle$ with two domain walls in Eq.~\eqref{domainex}.
(c) A ground state of the 2D kinetic Ising model with straight domain walls.
(d) One-dimensional gapless excitations on the surface of domain walls that corresponds to (b).
\label{kineticIsing}}
\end{figure}

To write down ground states and low-energy excitations of this model, let us introduce $|\bar{n}\rangle\coloneqq|\underbrace{1\cdots1}_{L-n}\underbrace{0\cdots0}_{n}\rangle$ in addition to $|n\rangle\coloneqq |\underbrace{0\cdots0}_{L-n}\underbrace{1\cdots1}_{n}\rangle$. 
For example, $|n\rangle$ satisfies
\begin{align}
&\hat{H}|n\rangle-2|n\rangle\notag\\
&=
-\frac{1}{2}(|n+1\rangle+|n-1\rangle+\hat{T}_1|n+1\rangle+\hat{T}_1^{-1}|n-1\rangle)
\end{align}
for $n=2,3,\cdots, L-2$, and
\begin{align}
&\hat{H}|n\rangle-2|n\rangle\notag\\
&=
\begin{cases}
-\frac{1}{2}(|2\rangle+\hat{T}_1|2\rangle)&(n=1),\\
-\frac{1}{2}(|L-2\rangle+\hat{T}_1^{-1}|L-2\rangle)&(n=L-1)
\end{cases}
\end{align}
for $n=1$ and $L-1$. Based on these expressions, the ground states under OBC are found to be
\begin{align}
|0\cdots0\rangle\,(=|\Phi_0\rangle), |1\cdots1\rangle, \sum_{n=1}^{L-1}|n\rangle,  \sum_{n=1}^{L-1}|\bar{n}\rangle.\label{GSGHZ}
\end{align}
The latter two are not consistent with the PBC and are hence not ground states of $\hat{H}$.

Although the fully-polarized state $|{\Phi_0}\rangle$ [Fig~\ref{kineticIsing}(a)] is one of the ground states, the plane-wave state in Eq.~\eqref{BGvs} is not a low-energy state because $\langle\Psi_k|\hat{H}|{\Psi_k}\rangle=2$. We instead consider a plane-wave state of two domain walls illustrated in Fig~\ref{kineticIsing}(b):
\begin{align}
|\Psi_{k,\ell}\rangle\coloneqq\frac{\sqrt{2}}{L}\sum_{n=1}^{L-1}\sum_{m=0}^{L-1}e^{ikm}e^{-ikn/2}\sin(\tfrac{\pi n\ell}{L})\hat{T}_1^m|n\rangle\label{domainex}
\end{align}
for $\ell=1,2,\cdots,L-1$. We find that $|\Psi_{k,\ell}\rangle$ is an exact eigenstate of $\hat{H}$ and $\hat{T}_1$ with the energy eigenvalue 
\begin{align}
E_\ell(k)=4\sin^2\Big(\frac{\pi \ell}{2L}\Big)+4\cos\Big(\frac{\pi \ell}{L}\Big)\sin^2\Big(\frac{k}{4}\Big).\label{domainexE}
\end{align}
The first term is $O(L^{-2})$, which vanishes in the thermodynamic limit. The second term is $O(k^2)$.

Despite the presence of gapless excitations, all the correlation functions of the form
\begin{align}
\langle\Phi_0|\hat{\mathcal{O}}_{\bm{x}}^\dagger(\hat{\mathbbm{1}}-\hat{G})\hat{\mathcal{O}}'_{\bm{y}}|\Phi_0\rangle\label{GHZdecay}
 \end{align}
decay exponentially with the system size for any choice of local operators $\hat{\mathcal{O}}_{\bm{x}}$ and $\hat{\mathcal{O}}'_{\bm{y}}$. Here $\hat G$ is the projector onto the ground states of $\hat{H}$. This is because both of the two ground states of $\hat{H}$ (the first two states in Eq.~\eqref{GSGHZ}) are product states.

\subsubsection{2D}
In two dimensions, the critical point $J_c$ is known to be $(1/2)\log(1+\sqrt{2})=0.440687\cdots$.
Above $J>J_c$, the $\mathbb{Z}_2$ symmetry is spontaneously broken and many quasi-degenerate ground states appear, whose finite size splitting decay exponentially with the system size. Hence one may think that this model is a counterexample to our first conjecture that states the absence of finite size splitting in gapped FF system. However, it is not the case because the model is known to be gapless above $J>J_c$~\cite{miyashitaDynamicalNaturePhase1985,henkelNonEquilibriumPhaseTransitions2010}. For example, in the limit $J\to \infty$, not only all spin up state and all spin down states, there are many other ground states with straight domain walls as illustrated in Fig~\ref{kineticIsing}(c)~\footnote{The width of each domain must be greater than one}.  The one-dimensional excitation on the surface of a domain wall [Fig~\ref{kineticIsing}(d)] that precisely takes the form $|\Psi_{k,\ell}\rangle$ in Eq.~\eqref{domainex} is a gapless excitation with the same excitation energy $E_\ell(k)$ in Eq.~\eqref{domainexE}.

\subsection{XYZ MG model}
As another nontrivial example, let us discuss a variation of the MG model~\cite{SRIRAMSHASTRY19811069,PhysRevLett.47.964,PhysRevB.57.11504,MG2,PhysRevB.103.214428}. The local Hamiltonian of the original model is given by
\begin{align}
&\hat{Q}_{x,x+1,x+2}\coloneqq\frac{1}{2}\hat{\mathbbm{1}}+
\frac{2}{3}(\hat{\bm{s}}_x\cdot\hat{\bm{s}}_{x+1}+\hat{\bm{s}}_{x+1}\cdot\hat{\bm{s}}_{x+2}+\hat{\bm{s}}_x\cdot\hat{\bm{s}}_{x+2}).
\end{align}
Any singlet state $|s\rangle_{x,x+1}\coloneqq(|01\rangle-|10\rangle)/\sqrt{2}$ belongs to the kernel of $\hat{Q}_{x,x+1,x+2}$:
\begin{align}
&\mathrm{span}\Big\{|010\rangle-|001\rangle,|100\rangle-|010\rangle,\notag\\
&\quad\quad\quad|110\rangle-|101\rangle,|101\rangle-|011\rangle\Big\}.
\end{align}

The extension of this model to the XYZ coupling
\begin{align}
\hat{Q}_{x,x+1,x+2}'\coloneqq \frac{1}{4}\hat{\mathbbm{1}}+\sum_{a=x,y,z}J_a(\hat{s}_x^a\hat{s}_{x+1}^a+\hat{s}_{x+1}^a\hat{s}_{x+2}^a+\hat{s}_x^a\hat{s}_{x+2}^a)\label{HXYZ}
\end{align}
is still frustration free for some ranges of $J_x$, $J_y$, $J_z$~\cite{SRIRAMSHASTRY19811069,PhysRevLett.47.964,PhysRevB.57.11504,MG2,PhysRevB.103.214428}. In particular, when
\begin{align}
J_x J_y+J_y J_z+J_z J_x=0,\label{MGXYZcondition}
\end{align}
the model exhibits additional ground state degeneracy. Following Ref.~\onlinecite{PhysRevB.103.214428}, we set 
\begin{align}
J_x&=\frac{1}{3}-\frac{2}{3}\cos\Big(2\theta-\frac{2\pi}{3}\Big),\\
J_y&=\frac{1}{3}-\frac{2}{3}\cos\Big(2\theta+\frac{2\pi}{3}\Big),\\
J_z&=\frac{1}{3}-\frac{2}{3}\cos2\theta\label{MGXYZcondition2}
\end{align}
with $-\pi/2\leq\theta\leq\pi/2$ so that the condition in Eq.~\eqref{MGXYZcondition} is satisfied and $\hat{Q}_{x,x+1,x+2}'$ becomes a projector. The kernel of $\hat{Q}_{x,x+1,x+2}'$ is given by
\begin{align}
&\mathrm{ker}\,\hat{Q}_{x,x+1,x+2}'=\mathrm{ker}\,\hat{Q}_{x,x+1,x+2}\oplus\mathrm{span}\big\{|\phi\rangle,|\bar{\phi}\rangle \big\},
\end{align}
where~\footnote{The presence of $|\bar{\phi}\rangle$ in the kernel is not relevant. Hence, the rank of $\hat{Q}_{x,x+1,x+2}$ can be three rather than two.}
\begin{align}
|\phi\rangle=\cos\theta\,|000\rangle+\frac{\sin\theta}{\sqrt{3}}(|011\rangle+|101\rangle+|110\rangle),\\
|\bar{\phi}\rangle=\cos\theta\,|111\rangle+\frac{\sin\theta}{\sqrt{3}}(|100\rangle+|010\rangle+|001\rangle).
\end{align}
Unless $\theta=0$, the fully polarized state $|{\Phi_0}\rangle$ is not a ground state of the system. 
The $\theta=\pm\pi/4$ cases are unitary equivalent to the model discussed in Ref.~\onlinecite{SaitoHotta}. 
The plane-wave state $|{\Psi_k}\rangle$ in Eq.~(4) of the main text is not a low-energy state unless $\theta=0$; indeed, we find
\begin{align}
\langle\Psi_k|\hat{H}|{\Psi_k}\rangle=\frac{\cos^2\theta}{3}(1+2 \cos k)^2+(L - 3)\sin^2\theta.
\end{align}

To construct a low-energy excitation, we assume $L$ is an odd integer greater than three. 
We introduce a plane-wave state of the domain wall state
\begin{align}
&|\tilde{\Psi}_k\rangle=\frac{1}{\sqrt{Ln_k}}\sum_{m=0}^{L-1}e^{ikm}\hat{T}_1^m|D\rangle,\\
&|D\rangle\coloneqq|\phi\rangle_{1,2,3}|s\rangle_{4,5}|s\rangle_{6,7}\cdots|s\rangle_{L-1,L},
\end{align}
where $n_k\coloneqq1-(-2)^{-\frac{L-5}{2}}\cos k$ is the normalization factor that converges to $1$ in the large $L$ limit. Using the translation symmetry of $\hat{H}$, we find
\begin{align}
\langle\tilde{\Psi}_k|\hat{H}|\tilde{\Psi}_k\rangle&=\frac{\big\|\hat{Q}_{2,3,4}(|D\rangle+\hat{T}_1^2|D\rangle)\big\|^2}{n_k}=\frac{2\sin^2k}{3n_k}.
\end{align}
This is $O(k^2)$, which is consistent with the result of Ref.~\onlinecite{PhysRevB.103.214428} on Kagome lattice. Hence, our conjecture still holds in this example as well, despite the fact that $|{\Phi_0}\rangle$ is not a ground state and $|{\Psi_k}\rangle$ is not gapless.

\section{Conclusion}
In this work, we proposed two conjectures regarding FF systems.  The first one posited that no finite-size splitting opens between the degenerate ground states. While this has been implicitly assumed in previous studies~\cite{BravyiGosset,Gossetozgunov,GossetHuang}, we pointed out the necessity of being cautious as it is actually a non-trivial assumption. The second conjecture stated that in a gapless FF system with translational symmetry, the dispersion relation of low-energy excitations near a certain wave number $\bm{k}_0$ can be bounded above by a quadratic dispersion. We proved this for the case of $s=1/2$ spin models with nearest-neighbor interactions defined on the hyper cubic lattice. However, this method could not be directly extended to models with longer-range interactions or larger local Hilbert space dimensions. Extending our results to these cases constitutes important future work.

Finally, there is a related, more general conjecture that
the finite-size gap $\epsilon\coloneqq E_{\tilde{N}_{\mathrm{deg}}+1}$ for gapless FF Hamiltonians can be bounded above by $O(L^{-2})$ regardless of the presence or absence of translational symmetry or the details of the boundary conditions.  In fact, in the accompanying paper~\cite{masaoka}, we prove this statement for critical FF systems in which a ground-state correlation function shows a power-law behavior. However, such an argument is not applicable to the uncle Hamiltonian for the GHZ state since all correlation functions decays exponentially.

\begin{acknowledgments}
We thank Hosho Katsura and Zijian Wang for useful discussions.
The work of H.W. is supported by JSPS KAKENHI Grant No.~JP24K00541.
This research is funded in part by the
Gordon and Betty Moore Foundation’s EPiQS Initiative,
Grant GBMF8683 to T.S.
\end{acknowledgments}

\bibliography{refs}

\begin{thebibliography}{30}%
\makeatletter
\providecommand \@ifxundefined [1]{%
 \@ifx{#1\undefined}
}%
\providecommand \@ifnum [1]{%
 \ifnum #1\expandafter \@firstoftwo
 \else \expandafter \@secondoftwo
 \fi
}%
\providecommand \@ifx [1]{%
 \ifx #1\expandafter \@firstoftwo
 \else \expandafter \@secondoftwo
 \fi
}%
\providecommand \natexlab [1]{#1}%
\providecommand \enquote  [1]{``#1''}%
\providecommand \bibnamefont  [1]{#1}%
\providecommand \bibfnamefont [1]{#1}%
\providecommand \citenamefont [1]{#1}%
\providecommand \href@noop [0]{\@secondoftwo}%
\providecommand \href [0]{\begingroup \@sanitize@url \@href}%
\providecommand \@href[1]{\@@startlink{#1}\@@href}%
\providecommand \@@href[1]{\endgroup#1\@@endlink}%
\providecommand \@sanitize@url [0]{\catcode `\\12\catcode `\$12\catcode
  `\&12\catcode `\#12\catcode `\^12\catcode `\_12\catcode `\%12\relax}%
\providecommand \@@startlink[1]{}%
\providecommand \@@endlink[0]{}%
\providecommand \url  [0]{\begingroup\@sanitize@url \@url }%
\providecommand \@url [1]{\endgroup\@href {#1}{\urlprefix }}%
\providecommand \urlprefix  [0]{URL }%
\providecommand \Eprint [0]{\href }%
\providecommand \doibase [0]{http://dx.doi.org/}%
\providecommand \selectlanguage [0]{\@gobble}%
\providecommand \bibinfo  [0]{\@secondoftwo}%
\providecommand \bibfield  [0]{\@secondoftwo}%
\providecommand \translation [1]{[#1]}%
\providecommand \BibitemOpen [0]{}%
\providecommand \bibitemStop [0]{}%
\providecommand \bibitemNoStop [0]{.\EOS\space}%
\providecommand \EOS [0]{\spacefactor3000\relax}%
\providecommand \BibitemShut  [1]{\csname bibitem#1\endcsname}%
\let\auto@bib@innerbib\@empty
\bibitem [{\citenamefont {Auerbach}(2012)}]{Auerbach}%
  \BibitemOpen
  \bibfield  {author} {\bibinfo {author} {\bibfnamefont {Assa}\ \bibnamefont
  {Auerbach}},\ }\href@noop {} {\emph {\bibinfo {title} {Interacting electrons
  and quantum magnetism}}}\ (\bibinfo  {publisher} {Springer Science \&
  Business Media},\ \bibinfo {year} {2012})\BibitemShut {NoStop}%
\bibitem [{\citenamefont {Tasaki}(2020)}]{TasakiBook}%
  \BibitemOpen
  \bibfield  {author} {\bibinfo {author} {\bibfnamefont {Hal}\ \bibnamefont
  {Tasaki}},\ }\href@noop {} {\emph {\bibinfo {title} {Physics and Mathematics
  of Quantum Many-Body Systems}}}\ (\bibinfo  {publisher} {Springer},\ \bibinfo
  {year} {2020})\BibitemShut {NoStop}%
\bibitem [{\citenamefont {Fannes}\ \emph {et~al.}(1992)\citenamefont {Fannes},
  \citenamefont {Nachtergaele},\ and\ \citenamefont {Werner}}]{parent}%
  \BibitemOpen
  \bibfield  {author} {\bibinfo {author} {\bibfnamefont {M.}~\bibnamefont
  {Fannes}}, \bibinfo {author} {\bibfnamefont {B.}~\bibnamefont
  {Nachtergaele}}, \ and\ \bibinfo {author} {\bibfnamefont {R.~F.}\
  \bibnamefont {Werner}},\ }\bibfield  {title} {\enquote {\bibinfo {title}
  {Finitely correlated states on quantum spin chains},}\ }\href {\doibase
  10.1007/BF02099178} {\bibfield  {journal} {\bibinfo  {journal}
  {Communications in Mathematical Physics}\ }\textbf {\bibinfo {volume}
  {144}},\ \bibinfo {pages} {443--490} (\bibinfo {year} {1992})}\BibitemShut
  {NoStop}%
\bibitem [{\citenamefont {Fern{\'a}ndez-Gonz{\'a}lez}\ \emph
  {et~al.}(2015)\citenamefont {Fern{\'a}ndez-Gonz{\'a}lez}, \citenamefont
  {Schuch}, \citenamefont {Wolf}, \citenamefont {Cirac},\ and\ \citenamefont
  {P{\'e}rez-Garc{\'\i}a}}]{uncle}%
  \BibitemOpen
  \bibfield  {author} {\bibinfo {author} {\bibfnamefont {C.}~\bibnamefont
  {Fern{\'a}ndez-Gonz{\'a}lez}}, \bibinfo {author} {\bibfnamefont
  {N.}~\bibnamefont {Schuch}}, \bibinfo {author} {\bibfnamefont {M.~M.}\
  \bibnamefont {Wolf}}, \bibinfo {author} {\bibfnamefont {J.~I.}\ \bibnamefont
  {Cirac}}, \ and\ \bibinfo {author} {\bibfnamefont {D.}~\bibnamefont
  {P{\'e}rez-Garc{\'\i}a}},\ }\bibfield  {title} {\enquote {\bibinfo {title}
  {Frustration free gapless hamiltonians for matrix product states},}\ }\href
  {\doibase 10.1007/s00220-014-2173-z} {\bibfield  {journal} {\bibinfo
  {journal} {Communications in Mathematical Physics}\ }\textbf {\bibinfo
  {volume} {333}},\ \bibinfo {pages} {299--333} (\bibinfo {year}
  {2015})}\BibitemShut {NoStop}%
\bibitem [{\citenamefont {Ogunnaike}\ \emph {et~al.}(2023)\citenamefont
  {Ogunnaike}, \citenamefont {Feldmeier},\ and\ \citenamefont
  {Lee}}]{PhysRevLett.131.220403}%
  \BibitemOpen
  \bibfield  {author} {\bibinfo {author} {\bibfnamefont {Olumakinde}\
  \bibnamefont {Ogunnaike}}, \bibinfo {author} {\bibfnamefont {Johannes}\
  \bibnamefont {Feldmeier}}, \ and\ \bibinfo {author} {\bibfnamefont
  {Jong~Yeon}\ \bibnamefont {Lee}},\ }\bibfield  {title} {\enquote {\bibinfo
  {title} {Unifying emergent hydrodynamics and lindbladian low-energy spectra
  across symmetries, constraints, and long-range interactions},}\ }\href
  {\doibase 10.1103/PhysRevLett.131.220403} {\bibfield  {journal} {\bibinfo
  {journal} {Phys. Rev. Lett.}\ }\textbf {\bibinfo {volume} {131}},\ \bibinfo
  {pages} {220403} (\bibinfo {year} {2023})}\BibitemShut {NoStop}%
\bibitem [{\citenamefont {Watanabe}\ \emph {et~al.}(2023)\citenamefont
  {Watanabe}, \citenamefont {Katsura},\ and\ \citenamefont
  {Lee}}]{arXiv:2310.16881}%
  \BibitemOpen
  \bibfield  {author} {\bibinfo {author} {\bibfnamefont {Haruki}\ \bibnamefont
  {Watanabe}}, \bibinfo {author} {\bibfnamefont {Hosho}\ \bibnamefont
  {Katsura}}, \ and\ \bibinfo {author} {\bibfnamefont {Jong~Yeon}\ \bibnamefont
  {Lee}},\ }\href@noop {} {\enquote {\bibinfo {title} {Spontaneous breaking of
  u(1) symmetry at zero temperature in one dimension},}\ } (\bibinfo {year}
  {2023}),\ \Eprint {http://arxiv.org/abs/2310.16881} {arXiv:2310.16881}
  \BibitemShut {NoStop}%
\bibitem [{\citenamefont {Palle}\ and\ \citenamefont
  {Benton}(2021)}]{PhysRevB.103.214428}%
  \BibitemOpen
  \bibfield  {author} {\bibinfo {author} {\bibfnamefont {Grgur}\ \bibnamefont
  {Palle}}\ and\ \bibinfo {author} {\bibfnamefont {Owen}\ \bibnamefont
  {Benton}},\ }\bibfield  {title} {\enquote {\bibinfo {title} {Exactly solvable
  spin-$\frac{1}{2}$ xyz models with highly degenerate partially ordered ground
  states},}\ }\href {\doibase 10.1103/PhysRevB.103.214428} {\bibfield
  {journal} {\bibinfo  {journal} {Phys. Rev. B}\ }\textbf {\bibinfo {volume}
  {103}},\ \bibinfo {pages} {214428} (\bibinfo {year} {2021})}\BibitemShut
  {NoStop}%
\bibitem [{\citenamefont {Ren}\ \emph {et~al.}(2024)\citenamefont {Ren},
  \citenamefont {Wang},\ and\ \citenamefont {Fang}}]{arXiv:2405.00785}%
  \BibitemOpen
  \bibfield  {author} {\bibinfo {author} {\bibfnamefont {Jie}\ \bibnamefont
  {Ren}}, \bibinfo {author} {\bibfnamefont {Yu-Peng}\ \bibnamefont {Wang}}, \
  and\ \bibinfo {author} {\bibfnamefont {Chen}\ \bibnamefont {Fang}},\
  }\href@noop {} {\enquote {\bibinfo {title} {Quasi-nambu-goldstone modes in
  many-body scar models},}\ } (\bibinfo {year} {2024}),\ \Eprint
  {http://arxiv.org/abs/2405.00785} {arXiv:2405.00785} \BibitemShut {NoStop}%
\bibitem [{\citenamefont {Bravyi}\ and\ \citenamefont
  {Gosset}(2015)}]{BravyiGosset}%
  \BibitemOpen
  \bibfield  {author} {\bibinfo {author} {\bibfnamefont {Sergey}\ \bibnamefont
  {Bravyi}}\ and\ \bibinfo {author} {\bibfnamefont {David}\ \bibnamefont
  {Gosset}},\ }\bibfield  {title} {\enquote {\bibinfo {title} {{Gapped and
  gapless phases of frustration-free spin- 1 2 chains}},}\ }\href {\doibase
  10.1063/1.4922508} {\bibfield  {journal} {\bibinfo  {journal} {J. Math.
  Phys.}\ }\textbf {\bibinfo {volume} {56}},\ \bibinfo {pages} {061902}
  (\bibinfo {year} {2015})}\BibitemShut {NoStop}%
\bibitem [{\citenamefont {Gosset}\ and\ \citenamefont
  {Mozgunov}(2016)}]{Gossetozgunov}%
  \BibitemOpen
  \bibfield  {author} {\bibinfo {author} {\bibfnamefont {David}\ \bibnamefont
  {Gosset}}\ and\ \bibinfo {author} {\bibfnamefont {Evgeny}\ \bibnamefont
  {Mozgunov}},\ }\bibfield  {title} {\enquote {\bibinfo {title} {{Local gap
  threshold for frustration-free spin systems}},}\ }\href {\doibase
  10.1063/1.4962337} {\bibfield  {journal} {\bibinfo  {journal} {Journal of
  Mathematical Physics}\ }\textbf {\bibinfo {volume} {57}},\ \bibinfo {pages}
  {091901} (\bibinfo {year} {2016})}\BibitemShut {NoStop}%
\bibitem [{\citenamefont {Gosset}\ and\ \citenamefont
  {Huang}(2016)}]{GossetHuang}%
  \BibitemOpen
  \bibfield  {author} {\bibinfo {author} {\bibfnamefont {David}\ \bibnamefont
  {Gosset}}\ and\ \bibinfo {author} {\bibfnamefont {Yichen}\ \bibnamefont
  {Huang}},\ }\bibfield  {title} {\enquote {\bibinfo {title} {Correlation
  length versus gap in frustration-free systems},}\ }\href {\doibase
  10.1103/PhysRevLett.116.097202} {\bibfield  {journal} {\bibinfo  {journal}
  {Phys. Rev. Lett.}\ }\textbf {\bibinfo {volume} {116}},\ \bibinfo {pages}
  {097202} (\bibinfo {year} {2016})}\BibitemShut {NoStop}%
\bibitem [{\citenamefont {Knabe}(1988)}]{Knabe}%
  \BibitemOpen
  \bibfield  {author} {\bibinfo {author} {\bibfnamefont {Stefan}\ \bibnamefont
  {Knabe}},\ }\bibfield  {title} {\enquote {\bibinfo {title} {Energy gaps and
  elementary excitations for certain vbs-quantum antiferromagnets},}\ }\href
  {\doibase 10.1007/BF01019721} {\bibfield  {journal} {\bibinfo  {journal} {J.
  Stat. Phys.}\ }\textbf {\bibinfo {volume} {52}},\ \bibinfo {pages} {627--638}
  (\bibinfo {year} {1988})}\BibitemShut {NoStop}%
\bibitem [{\citenamefont {Anshu}(2020)}]{Anshu}%
  \BibitemOpen
  \bibfield  {author} {\bibinfo {author} {\bibfnamefont {Anurag}\ \bibnamefont
  {Anshu}},\ }\bibfield  {title} {\enquote {\bibinfo {title} {Improved local
  spectral gap thresholds for lattices of finite size},}\ }\href {\doibase
  10.1103/PhysRevB.101.165104} {\bibfield  {journal} {\bibinfo  {journal}
  {Phys. Rev. B}\ }\textbf {\bibinfo {volume} {101}},\ \bibinfo {pages}
  {165104} (\bibinfo {year} {2020})}\BibitemShut {NoStop}%
\bibitem [{\citenamefont {Lemm}\ and\ \citenamefont {Xiang}(2022)}]{Lemm_2022}%
  \BibitemOpen
  \bibfield  {author} {\bibinfo {author} {\bibfnamefont {Marius}\ \bibnamefont
  {Lemm}}\ and\ \bibinfo {author} {\bibfnamefont {David}\ \bibnamefont
  {Xiang}},\ }\bibfield  {title} {\enquote {\bibinfo {title} {Quantitatively
  improved finite-size criteria for spectral gaps},}\ }\href {\doibase
  10.1088/1751-8121/ac7989} {\bibfield  {journal} {\bibinfo  {journal} {Journal
  of Physics A: Mathematical and Theoretical}\ }\textbf {\bibinfo {volume}
  {55}},\ \bibinfo {pages} {295203} (\bibinfo {year} {2022})}\BibitemShut
  {NoStop}%
\bibitem [{\citenamefont {Katsura}\ \emph {et~al.}(2015)\citenamefont
  {Katsura}, \citenamefont {Schuricht},\ and\ \citenamefont
  {Takahashi}}]{PhysRevB.92.115137}%
  \BibitemOpen
  \bibfield  {author} {\bibinfo {author} {\bibfnamefont {Hosho}\ \bibnamefont
  {Katsura}}, \bibinfo {author} {\bibfnamefont {Dirk}\ \bibnamefont
  {Schuricht}}, \ and\ \bibinfo {author} {\bibfnamefont {Masahiro}\
  \bibnamefont {Takahashi}},\ }\bibfield  {title} {\enquote {\bibinfo {title}
  {Exact ground states and topological order in interacting kitaev/majorana
  chains},}\ }\href {\doibase 10.1103/PhysRevB.92.115137} {\bibfield  {journal}
  {\bibinfo  {journal} {Phys. Rev. B}\ }\textbf {\bibinfo {volume} {92}},\
  \bibinfo {pages} {115137} (\bibinfo {year} {2015})}\BibitemShut {NoStop}%
\bibitem [{\citenamefont {Wouters}\ \emph {et~al.}(2018)\citenamefont
  {Wouters}, \citenamefont {Katsura},\ and\ \citenamefont
  {Schuricht}}]{PhysRevB.98.155119}%
  \BibitemOpen
  \bibfield  {author} {\bibinfo {author} {\bibfnamefont {Jurriaan}\
  \bibnamefont {Wouters}}, \bibinfo {author} {\bibfnamefont {Hosho}\
  \bibnamefont {Katsura}}, \ and\ \bibinfo {author} {\bibfnamefont {Dirk}\
  \bibnamefont {Schuricht}},\ }\bibfield  {title} {\enquote {\bibinfo {title}
  {Exact ground states for interacting kitaev chains},}\ }\href {\doibase
  10.1103/PhysRevB.98.155119} {\bibfield  {journal} {\bibinfo  {journal} {Phys.
  Rev. B}\ }\textbf {\bibinfo {volume} {98}},\ \bibinfo {pages} {155119}
  (\bibinfo {year} {2018})}\BibitemShut {NoStop}%
\bibitem [{\citenamefont {Witten}(1982)}]{WITTEN1982253}%
  \BibitemOpen
  \bibfield  {author} {\bibinfo {author} {\bibfnamefont {Edward}\ \bibnamefont
  {Witten}},\ }\bibfield  {title} {\enquote {\bibinfo {title} {Constraints on
  supersymmetry breaking},}\ }\href {\doibase
  https://doi.org/10.1016/0550-3213(82)90071-2} {\bibfield  {journal} {\bibinfo
   {journal} {Nuclear Physics B}\ }\textbf {\bibinfo {volume} {202}},\ \bibinfo
  {pages} {253--316} (\bibinfo {year} {1982})}\BibitemShut {NoStop}%
\bibitem [{\citenamefont {Wouters}\ \emph {et~al.}(2021)\citenamefont
  {Wouters}, \citenamefont {Katsura},\ and\ \citenamefont
  {Schuricht}}]{katsura}%
  \BibitemOpen
  \bibfield  {author} {\bibinfo {author} {\bibfnamefont {Jurriaan}\
  \bibnamefont {Wouters}}, \bibinfo {author} {\bibfnamefont {Hosho}\
  \bibnamefont {Katsura}}, \ and\ \bibinfo {author} {\bibfnamefont {Dirk}\
  \bibnamefont {Schuricht}},\ }\bibfield  {title} {\enquote {\bibinfo {title}
  {{Interrelations among frustration-free models via Witten's conjugation}},}\
  }\href {\doibase 10.21468/SciPostPhysCore.4.4.027} {\bibfield  {journal}
  {\bibinfo  {journal} {SciPost Phys. Core}\ }\textbf {\bibinfo {volume} {4}},\
  \bibinfo {pages} {027} (\bibinfo {year} {2021})}\BibitemShut {NoStop}%
\bibitem [{\citenamefont {{Wikipedia article}}()}]{Lucas}%
  \BibitemOpen
  \bibfield  {author} {\bibinfo {author} {\bibnamefont {{Wikipedia article}}},\
  }\href {https://en.wikipedia.org/wiki/Lucas_number} {\enquote {\bibinfo
  {title} {Lucas number},}\ }\BibitemShut {NoStop}%
\bibitem [{Note1()}]{Note1}%
  \BibitemOpen
  \bibinfo {note} {This argument cannot be directly applied to the simpler
  decomposition $\protect \hat {H}^{(1)}+\protect \hat {H}^{(2)}$, because if
  $N_{\protect \mathrm {deg}}\geq 2$ for each $\protect \hat {H}_y^{\protect
  \mathrm {1D}}$, then the ground state degeneracy of $\protect \hat {H}^{(1)}$
  becomes $N_{\protect \mathrm {deg}}^L$, which grows exponentially with the
  system size and the min-max principle becomes silent. For the same reason,
  the case of $\alpha =\beta =\gamma =0$ in the rank 1 model should be treated
  separately, but this case is trivially gapped in any spatial
  dimension.}\BibitemShut {Stop}%
\bibitem [{\citenamefont {Masaoka}\ \emph {et~al.}(2024)\citenamefont
  {Masaoka}, \citenamefont {Soejima},\ and\ \citenamefont
  {Watanabe}}]{masaoka}%
  \BibitemOpen
  \bibfield  {author} {\bibinfo {author} {\bibfnamefont {Rintaro}\ \bibnamefont
  {Masaoka}}, \bibinfo {author} {\bibfnamefont {Tomohiro}\ \bibnamefont
  {Soejima}}, \ and\ \bibinfo {author} {\bibfnamefont {Haruki}\ \bibnamefont
  {Watanabe}},\ }\href@noop {} {} (\bibinfo {year} {2024}),\ \Eprint
  {http://arxiv.org/abs/2406.06415} {arXiv:2406.06415} \BibitemShut {NoStop}%
\bibitem [{\citenamefont {Miyashita}\ and\ \citenamefont
  {Takano}(1985)}]{miyashitaDynamicalNaturePhase1985}%
  \BibitemOpen
  \bibfield  {author} {\bibinfo {author} {\bibfnamefont {Seiji}\ \bibnamefont
  {Miyashita}}\ and\ \bibinfo {author} {\bibfnamefont {Hiroshi}\ \bibnamefont
  {Takano}},\ }\bibfield  {title} {\enquote {\bibinfo {title} {Dynamical nature
  of the phase transition of the two-dimensional kinetic ising model},}\ }\href
  {\doibase 10.1143/PTP.73.1122} {\bibfield  {journal} {\bibinfo  {journal}
  {Progress of Theoretical Physics}\ }\textbf {\bibinfo {volume} {73}},\
  \bibinfo {pages} {1122--1140} (\bibinfo {year} {1985})}\BibitemShut {NoStop}%
\bibitem [{\citenamefont {Henkel}\ and\ \citenamefont
  {Pleimling}(2010)}]{henkelNonEquilibriumPhaseTransitions2010}%
  \BibitemOpen
  \bibfield  {author} {\bibinfo {author} {\bibfnamefont {Malte}\ \bibnamefont
  {Henkel}}\ and\ \bibinfo {author} {\bibfnamefont {Michel}\ \bibnamefont
  {Pleimling}},\ }\href {\doibase 10.1007/978-90-481-2869-3} {\emph {\bibinfo
  {title} {Non-Equilibrium Phase Transitions}}},\ Theoretical and Mathematical
  Physics\ (\bibinfo  {publisher} {Springer Netherlands},\ \bibinfo {address}
  {Dordrecht},\ \bibinfo {year} {2010})\BibitemShut {NoStop}%
\bibitem [{Note2()}]{Note2}%
  \BibitemOpen
  \bibinfo {note} {The width of each domain must be greater than
  one}\BibitemShut {NoStop}%
\bibitem [{\citenamefont {{Sriram Shastry}}\ and\ \citenamefont
  {Sutherland}(1981)}]{SRIRAMSHASTRY19811069}%
  \BibitemOpen
  \bibfield  {author} {\bibinfo {author} {\bibfnamefont {B.}~\bibnamefont
  {{Sriram Shastry}}}\ and\ \bibinfo {author} {\bibfnamefont {Bill}\
  \bibnamefont {Sutherland}},\ }\bibfield  {title} {\enquote {\bibinfo {title}
  {Exact ground state of a quantum mechanical antiferromagnet},}\ }\href
  {\doibase https://doi.org/10.1016/0378-4363(81)90838-X} {\bibfield  {journal}
  {\bibinfo  {journal} {Physica B+C}\ }\textbf {\bibinfo {volume} {108}},\
  \bibinfo {pages} {1069--1070} (\bibinfo {year} {1981})}\BibitemShut {NoStop}%
\bibitem [{\citenamefont {Shastry}\ and\ \citenamefont
  {Sutherland}(1981)}]{PhysRevLett.47.964}%
  \BibitemOpen
  \bibfield  {author} {\bibinfo {author} {\bibfnamefont {B.~Sriram}\
  \bibnamefont {Shastry}}\ and\ \bibinfo {author} {\bibfnamefont {Bill}\
  \bibnamefont {Sutherland}},\ }\bibfield  {title} {\enquote {\bibinfo {title}
  {Excitation spectrum of a dimerized next-neighbor antiferromagnetic chain},}\
  }\href {\doibase 10.1103/PhysRevLett.47.964} {\bibfield  {journal} {\bibinfo
  {journal} {Phys. Rev. Lett.}\ }\textbf {\bibinfo {volume} {47}},\ \bibinfo
  {pages} {964--967} (\bibinfo {year} {1981})}\BibitemShut {NoStop}%
\bibitem [{\citenamefont {Gerhardt}\ \emph {et~al.}(1998)\citenamefont
  {Gerhardt}, \citenamefont {M\"utter},\ and\ \citenamefont
  {Kr\"oger}}]{PhysRevB.57.11504}%
  \BibitemOpen
  \bibfield  {author} {\bibinfo {author} {\bibfnamefont {C.}~\bibnamefont
  {Gerhardt}}, \bibinfo {author} {\bibfnamefont {K.-H.}\ \bibnamefont
  {M\"utter}}, \ and\ \bibinfo {author} {\bibfnamefont {H.}~\bibnamefont
  {Kr\"oger}},\ }\bibfield  {title} {\enquote {\bibinfo {title} {Metamagnetism
  in the xxz model with next-to-nearest-neighbor coupling},}\ }\href {\doibase
  10.1103/PhysRevB.57.11504} {\bibfield  {journal} {\bibinfo  {journal} {Phys.
  Rev. B}\ }\textbf {\bibinfo {volume} {57}},\ \bibinfo {pages} {11504--11509}
  (\bibinfo {year} {1998})}\BibitemShut {NoStop}%
\bibitem [{\citenamefont {Xu}\ \emph {et~al.}(2021)\citenamefont {Xu},
  \citenamefont {Zhang}, \citenamefont {Guo},\ and\ \citenamefont
  {Gong}}]{MG2}%
  \BibitemOpen
  \bibfield  {author} {\bibinfo {author} {\bibfnamefont {Hong-Ze}\ \bibnamefont
  {Xu}}, \bibinfo {author} {\bibfnamefont {Shun-Yao}\ \bibnamefont {Zhang}},
  \bibinfo {author} {\bibfnamefont {Guang-Can}\ \bibnamefont {Guo}}, \ and\
  \bibinfo {author} {\bibfnamefont {Ming}\ \bibnamefont {Gong}},\ }\bibfield
  {title} {\enquote {\bibinfo {title} {Exact dimer phase with anisotropic
  interaction for one dimensional magnets},}\ }\href {\doibase
  10.1038/s41598-021-85483-0} {\bibfield  {journal} {\bibinfo  {journal}
  {Scientific Reports}\ }\textbf {\bibinfo {volume} {11}},\ \bibinfo {pages}
  {6462} (\bibinfo {year} {2021})}\BibitemShut {NoStop}%
\bibitem [{Note3()}]{Note3}%
  \BibitemOpen
  \bibinfo {note} {The presence of $|\protect \bar {\phi }\rangle $ in the
  kernel is not relevant. Hence, the rank of $\protect \hat {Q}_{x,x+1,x+2}$
  can be three rather than two.}\BibitemShut {Stop}%
\bibitem [{\citenamefont {Saito}\ and\ \citenamefont
  {Hotta}(2024)}]{SaitoHotta}%
  \BibitemOpen
  \bibfield  {author} {\bibinfo {author} {\bibfnamefont {Hidehiro}\
  \bibnamefont {Saito}}\ and\ \bibinfo {author} {\bibfnamefont {Chisa}\
  \bibnamefont {Hotta}},\ }\bibfield  {title} {\enquote {\bibinfo {title}
  {Exact matrix product states at the quantum lifshitz tricritical point in a
  spin-$1/2$ zigzag-chain antiferromagnet with anisotropic
  $\mathrm{\ensuremath{\Gamma}}$ term},}\ }\href {\doibase
  10.1103/PhysRevLett.132.166701} {\bibfield  {journal} {\bibinfo  {journal}
  {Phys. Rev. Lett.}\ }\textbf {\bibinfo {volume} {132}},\ \bibinfo {pages}
  {166701} (\bibinfo {year} {2024})}\BibitemShut {NoStop}%
\end{thebibliography}%

\appendix

\section{Proof of the min-max principle}
\label{appminimax}
Let $\mathcal{M}_j$ be an arbitrary $j$ dimensional subspace of the entire Hilbert space $\mathcal{H}_D$. 
We \emph{maximize} the energy expectation value $\langle\Phi|\hat{H}|\Phi\rangle$ by varying the normalized state $|\Phi\rangle$ belonging to $\mathcal{M}_j$.
We then \emph{minimize} the maximum value $\max_{|\Phi\rangle\in\mathcal{M}_j}\langle\Phi|\hat{H}|\Phi\rangle$ by varying the subspace $\mathcal{M}_j$ of $\mathcal{H}_D$.
By definition, the minimum value $E_j$ is achieved when $\mathcal{M}_j$ is spanned by the $j$ eigenvectors corresponding to the eigenvalues $E_1, E_2, \cdots, E_j$.
Hence, we obtain an expression 
\begin{align}
E_j=\min_{\mathcal{M}_j\subset\mathcal{H}_D}\Big(\max_{|\Phi\rangle\in\mathcal{M}_j}\langle\Phi|\hat{H}|\Phi\rangle\Big).
\end{align}
The same argument leads to
\begin{align}
E_j'=\min_{\mathcal{M}_j\subset\mathcal{H}_D}\Big(\max_{|\Phi\rangle\in\mathcal{M}_j}\langle\Phi|\hat{H}'|\Phi\rangle\Big).
\end{align}
Finally, by the definition of $\hat{V}=\hat{H}'-\hat{H}\geq0$, we have
\begin{align}
\langle\Phi|\hat{H}'|\Phi\rangle\geq\langle\Phi|\hat{H}|\Phi\rangle
\end{align}
for any state $|\Phi\rangle$ in $\mathcal{H}_D$.
In particular, if $\langle\Phi|\hat{H}|\Phi\rangle$ and $\langle\Phi|\hat{H}'|\Phi\rangle$ are maximized by $|\Phi\rangle_*\in\mathcal{M}_j$ and  $|\Phi\rangle_*'\in\mathcal{M}_j$, respectively, we have
\begin{align}
\max_{|\Phi\rangle\in\mathcal{M}_j}\langle\Phi|\hat{H}'|\Phi\rangle&=\langle\Phi_*'|\hat{H}'|\Phi_*'\rangle
\geq\langle\Phi_*|\hat{H}'|\Phi_*\rangle\notag\\
&
\geq\langle\Phi_*|\hat{H}|\Phi_*\rangle=\max_{|\Phi\rangle\in\mathcal{M}_j}\langle\Phi|\hat{H}|\Phi\rangle.
\end{align}
This relation holds for any $\mathcal{M}_j\subset\mathcal{H}_D$. Hence, we arrive at the theorem.

\section{Derivation of Eq.~\eqref{BGHamiltonian3}}
\label{derivationpsi}
Let us consider a local unitary that maps $|\psi\rangle$ to $|\psi'\rangle=\hat{U}\otimes\hat{U}|\psi\rangle$.
In general, $\hat{U}$ can be parametrized by $u_1,u_2\in\mathbb{C}$ ($|u_1|^2+|u_2|^2=1$) and $\theta\in[0,2\pi)$  as
\begin{align}
&\hat{U}|0\rangle=e^{i\theta}(u_1|0\rangle+u_2|1\rangle),\label{unitary1}\\
&\hat{U}|1\rangle=e^{i\theta}(-u_2^*|0\rangle+u_1^*|1\rangle).\label{unitary2}
\end{align}
The inverse of $\hat{U}$ reads
\begin{align}
&\hat{U}^\dagger|0\rangle=e^{-i\theta}(u_1^*|0\rangle-u_2|1\rangle),\\
&\hat{U}^\dagger|1\rangle=e^{-i\theta}(u_2^*|0\rangle+u_1|1\rangle).
\end{align}
Under this local unitary transformation, the matrix $m_\psi$ defined by
\begin{align}
&m_\psi\coloneqq\begin{pmatrix}
\langle\psi|01\rangle&\langle\psi|11\rangle\\
-\langle\psi|00\rangle&
-\langle\psi|10\rangle
\end{pmatrix}
\end{align}
is changed to
\begin{align}
m_{\psi'}&=\begin{pmatrix}
\langle\psi|\hat{U}^\dagger\otimes\hat{U}^\dagger|01\rangle&\langle\psi|\hat{U}^\dagger\otimes\hat{U}^\dagger|11\rangle\\
-\langle\psi|\hat{U}^\dagger\otimes\hat{U}^\dagger|00\rangle&-\langle\psi|\hat{U}^\dagger\otimes\hat{U}^\dagger|10\rangle
\end{pmatrix}\notag\\
&=
e^{-2i\theta}\begin{pmatrix}
u_1&-u_2^*\\
u_2&u_1^*
\end{pmatrix}
\begin{pmatrix}
\langle\psi|01\rangle&\langle\psi|11\rangle\\
-\langle\psi|00\rangle&
-\langle\psi|10\rangle
\end{pmatrix}
\begin{pmatrix}
u_1^*&u_2^*\\
-u_2&u_1
\end{pmatrix}\notag\\
&=e^{-2i\theta}Um_{\psi} U^\dagger.
\end{align}
Therefore, by choosing $U=\begin{pmatrix}
u_1&-u_2^*\\
u_2&u_1^*
\end{pmatrix}$ and $\theta$ properly, one gets the form
\begin{align}
&m_\psi=\begin{pmatrix}
\alpha-i\beta&\delta\\
0&-\alpha+i\gamma
\end{pmatrix}.\label{Tpsi}
\end{align}
which is equivalent to the parametrization in Eq.~(6) of the main text.

\section{Derivation of Eq.~\eqref{Lucus}}
\label{secLucus}
Here we discuss the case $\alpha=\beta=0$ in the rank 1 model. (The  $\alpha=\gamma=0$ case can be treated in the same way.) In this case, $|\psi\rangle$ is not entangled because it can be written as a product state
\begin{align}
& |\psi\rangle=|1\rangle\otimes|v^\perp\rangle,
\end{align}
where $|v^\perp\rangle=i\gamma|0\rangle+\delta |1\rangle$ with $\gamma^2+\delta^2=1$.
According to Ref.~\onlinecite{BravyiGosset}, excitations under OBC are gapped in this case, which implies that excitations under PBC are also gapped. 

When $\gamma\neq0$, orthogonal ground states under OBC are product states with a single domain-wall:
\begin{align}
& |\underbrace{0\cdots0}_{n-1}v^{\perp}\underbrace{v\cdots v}_{L-n}\rangle\quad (n=1,\cdots L),
\end{align}
where $|v\rangle=\delta|0\rangle+i\gamma |1\rangle$ is orthogonal to $|v^\perp\rangle$. 
Translation-invariant product states $|v\cdots v\rangle$ and $|0\cdots0\rangle$ are also ground states but only the former is orthogonal to the domain-wall states and the latter is not linearly independent. In contrast, under PBC, domain wall states violate the boundary term and the two translation-invariant product states give the ground states. 

On the other hand, the  $\alpha=\beta=\gamma=0$ case is exceptional. In this case $|\psi\rangle=\pm |11\rangle$ is a symmetric product state and any product state with no consecutive $1$'s is a ground state of $\hat{H}^{\mathrm{obc}}$ and $\hat{H}^{\mathrm{pbc}}$. Hence, 
\begin{align}
&N_{\mathrm{deg}}^{\mathrm{obc}}=\sum_{n=0}^{\lfloor \frac{L+1}{2} \rfloor}\binom{L-n+1}{n},\\
&N_{\mathrm{deg}}^{\mathrm{pbc}}=N_{\mathrm{deg}}^{\mathrm{obc}}-\sum_{n=2}^{\lfloor \frac{L+1}{2} \rfloor}\binom{L-n-1}{n-2},
\end{align}
where $\lfloor x\rfloor$ represents the greatest integer less than or equal to $x\in\mathbb{R}$. The second term in $N_{\mathrm{deg}}^{\mathrm{pbc}}$ represents the number of valid product states under OBC which start with $|1\rangle_{x=1}$ and end with $|1\rangle_{x=L}$. 
The excitation gap is exactly 1.

\end{document}